\documentclass{article}\textheight 9in\textwidth 6.5in
\begin{document}

 \newcommand\mb[1]{\mbox{\Large$\mathbf{#1}$}}
 \newcommand\lf{\mb\triangleleft} \newcommand\rt{\mb\triangleright}
 \newtheorem{thr}{Theorem} \newcommand\qed{\rule{1em}{1.5ex}}

 \title {\null\vspace{-4pc}
 Self-stabilization of Circular Arrays of Automata\thanks
 {{\em Theor. Comp. Sci.}, 235(1):143-144, 3/17/2000.}}

 \author {Leonid A.~Levin\thanks
 {Supported by NSF grants CCR-9610455, CCR-9820934.}\\
 Boston University\thanks {Computer Science department, 111 Cummington St.,
 Boston, MA 02215; (e-mail to Lnd@bu.edu)}} \date{} \maketitle

\vspace{1pc} 

As computing elements shrink closer to molecular size, the reliability
becomes a serious issue. John von Neumann proposed circuits which compute
reliably in the presence of noise. They cannot be realized, however, with
local connections on an Euclidean grid.
 Andrei Toom designed a planar grid of simple automata which stores one bit
reliably despite noise. Aligned in layers they were used by Peter Gacs and
John Reif to simulate a Turing Machine on a three-dimensional grid.
 The need to dispose of heat restricts computing chips to 2 dimensions,
which would make this method to depend on linear arrays of automata to
preserve each bit reliably. Such automata were believed impossible until
Kurdiumov and Gacs constructed them in a series of monumental works.
 Their complexity is enormous, which leaves a need for simpler solutions.

[Gacs, Kurdiumov, Levin, 78] proposed simple automata with 2 states
``\rt,\lf''. They are arranged in a linear array and change state when
opposed by both first and third neighbor from the sharp end.
 In an infinite array they are self-stabilizing: if all but a finite
minority of automata are in the same state, the minority states disappear.
Implicit in the paper was a stronger result that a sufficiently small
minority of states vanish even in a finite circular array (ring).
 I think this strengthening deserves to be made explicit which is the
purpose of the following note.

 \begin{thr}. Let $\alpha=1/x\approx1.7734$ where $5^x=2^x+1$. Let in a ring of
more than $3k^\alpha$ GKL automata all but $k$ of them start in the same state.
Then the minority states disappear within $3k^\alpha$ steps. \end{thr}

Assume, the \rt-s are in majority. We refer as intervals to maximal strings
of three or more {\rt}-s and as segments to strings with {\lf} at each end.
Let $S$ be a segment between \rt$^{2|S|+1}$ and \rt$^{|S|+2}$.
 It grows at most one cell per step to the left and never to the right.
 The left end of its leftmost pattern {\lf\lf} or {\lf\rt\rt\lf} moves
right each step until gone. Then $S$ shrinks at the right end by at least 3
cells per step. So, $2|S|$ steps erase any effect of $S$.
 We say $S$ is {\em killed} by the surrounding {\rt} intervals.
 Any segment $S$ is killed unless it has a {\lf} within either $|S|+1$
cells to the right or $2|S|$ to the left. We say $S$ is closed to this side
and break a ring into a binary hierarchy of {\em solid} segments, each of
which can be killed only as a whole: We start with single \lf-s and combine
into a higher solid any two solid neighbors closed toward each other.

Let $S$ be a counterexample to the Theorem with the fewest \lf-s.
 Its maximal solids must all be closed from one side (and so occupy at
least a third of $S$). This side must be left: If all are open to the
left, then the left neighbor of the longest interval is killed; if some
are closed to the left and some to the right, they combine further. Take
the solid $P$ next right to the longest on $S$ interval $p$. Expanding
left, $P$ sheds its {\lf\lf} and {\lf\rt\rt\lf} patterns and shrinks
moving left until it hits its left neighbor and is overrun. Its right
end moves left for at least $2|p|-|P|-2$ cells, extending the original
$\ge|P|+2$ cells of the interval at its right enough to kill its right
neighbor. Thus, $S$ can have only one maximal solid, closed on both
sides, and $|S|\le2|P|+1$.

It is left to prove that no solid with $k$ \lf-s has more than $(3k^\alpha
-1)/2$ cells. Consider a shortest solid which violates that. It consists of
two solids of $n$ and $nt-1\ge n$ cells respectively and an interval of
$nr\le n\min\{2,t\}$ cells. It has at least $((2n+1)/3)^x+((2nt-1)/3)^x\ge
(2n/3)^x(1+t^x)$ \lf-s. We must see that this is at least $((2n+2nr+2nt-1)
/3)^x < (2n/3)^x(1+r+t)^x$, {i.e.} that $1+t^x - (1+t+\min\{2,t\})^x\ge0$.
Since $t>1$, this function has its minimum $0$ at $t=2$. \qed

The factor $3$ can be improved and David Metcalf conjectured that the power
can be improved too, to the $1/\log$ of the golden ratio $\approx1.44$ .
 It certainly cannot be improved beyond that: The strings $S_0=S_1=\lf^3$,
$S_{i+2}=S_i\rt^{|S_i|-4}S_{i+1}$ with $|S_i|=2^{i-1}+2$ have $3$
Fibonacci$(i)$ \lf-s and kill all their \rt-s.

\pagebreak \end{document}